# Metasurface-Based Free-Space Multi-Port Beam Splitter with Arbitrary Power Ratio

*Tian Tian[1], Yuxuan Liao[1], Xue Feng[1]\*, Kaiyu Cui[1], Fang Liu[1], Wei Zhang[1], and Yidong Huang[1]\**

[1]Department of Electronic Engineering, Tsinghua University, Beijing 100084, China

\*Correspondence: x-feng@tsinghua.edu.cn; yidonghuang@tsinghua.edu.cn

**ABSTRACT:** A beam splitter (BS) is one of the most critical building blocks in optical systems. Despite various attempts of flat-type BSs to miniaturize the conventional cube BS reported, it remains a challenge to realize an ultrathin optical BS with multi-port output, non-uniform splitting ratio and steerable outgoing directions. Herein, we have demonstrated a free-space optical multi-port beam splitter (MPBS) based on a polarization-independent all-dielectric metasurface. By applying an optimized phase-pattern paradigm via a gradient-descent-based iterative algorithm to amorphous silicon ($\alpha$-Si) metasurfaces, we have prepared a variety of MPBS samples with arbitrarily predetermined output port number (2~7), power ratio and spatial distribution of output beams. The experimental results reveal that the fabricated MPBSs could achieve high total splitting efficiency (*TSE*, above 74.7%) and beam-splitting fidelity (*similarity*, above 78.4%) within the bandwidth of 100 nm (1500~1600 nm). We envision that such MPBS could provide fabulous flexibility for optical integrated system and diverse applications.

**INTRODUCTION**

A beam splitter (BS), which could distribute and combine the optical beam paths, serves as one of the most fundamental building blocks in a variety of optical systems. Typically, the most common BS divides the incident light beam into two output channels (denoted as 2-port BS). Such 2-port BS plays a significant role in diverse optical applications of interferometers,[1][2] spectroscopy,[3] optical communications,[4] etc. Previously, the conventional 2-port BS has been intensively investigated in both free-space (e.g., the cube BS[5]) and on-chip (e.g., y-branch waveguide,[6][7] directional coupler[8]) optical systems. Nevertheless, the cube BS is hard to be integrated on-chip since the optical path relies on the birefringence of the bulky prisms. As more compact alternatives, some flat-type BSs (e.g., grating BS,[9] dichroic BS[10]) have been recently demonstrated. Besides the traditional 2-port BSs, we noticed that for some specific applications, e.g., light detection and ranging (LiDAR),[11] beamforming networks,[12] quantum optics[13] and optical computing,[14][15] a multi-port beam splitter (MPBS) is highly desired. The functionality of the MPBS is to distribute the incident beam into $N$ ($N>2$) output channels with predetermined power ratio. Furthermore, for these applications, the operation wavelength and polarization should keep constant since the optical interference can be employed to perform some processing or calculation in optical domain. Such a MPBS is actually very difficult to achieve through traditional methods in both free-space and integrated systems since both the wavelength and polarization cannot be utilized to guide the optical beam. For instance, the widely used multimode interference (MMI) coupler only can divide the incident beam into $N$ output channels with uniform power

ratio.[16][17] Another scheme is the multi-port interferometer, which is constructed by a specific mesh of 2×2 BSs and phase shifters.[18] Since such *N*-port interferometer requires cascaded $N\times(N-1)$ 2-port BSs, it is still challenging to fabricate and not cost-efficient. One promising method to achieve an optical MPBS with a single compact element is to employ the burgeoning metasurface.

Metasurface is a two-dimension artificial structure with periodic arrangement of subwavelength scattering unit cells, which are referred to as "meta-atoms". During the past decades, metasurface has attracted extensive attentions owing to its overwhelming superiority over conventional optical component.[19][20] Specifically, as a compact planar structure, metasurface meets the incremental pursuit of photonic integrated systems to miniaturize the footprint. Furthermore, with different materials (e.g., dieletric,[21][22] metal[23][24]) and geometric structure of unit cells,[25][26] it is flexible to design metasurface-empowered devices with the operation frequency band ranging from visible,[27][29] terahertz,[30] to microwave.[31] Notably, metasurface also exhibits the unprecedented capability to manipulate the electromagnetic wave in various degrees of freedom including phase,[32][34] amplitude[35][36] and polarization.[37] By taking advantage of these intriguing properties, metasurface could extend the functionalities of existing optical components[38][39] and further achieve novel devices such as perfect absorbers,[40] invisibility skin cloak,[41] Meta-Hologram,[42] Laplace metasurface,[43] etc. Based on metasurfaces, several optical 2-port BSs[44]-[48] have recently been demonstrated with ultra-compact size (33.6μm × 33.6μm as a concrete example[44]), which are very promising to replace the conventional cube BSs. In these work, only a few theoretical proposals realized the asymmetric (i.e., non-identical splitting angle or non-uniform power ratio) BSs based on the difference of incident angle[45] or frequency,[48] even by altering the refractive index of the substrate.[44] As mentioned above, for our considering MPBS, the wavelength and polarization cannot be utilized to separate the light beam as well as the incident angle since the incident beam is constant. Thus, the metasurface is the most promising candidate through the controllable transmission phase distribution within the transverse plane.

In this work, we have proposed and demonstrated a free-space optical MPBS with arbitrarily predetermined port number, power ratio and spatial distribution of output beams for a single wavelength input beam. The MPBS is fabricated on the amorphous silicon ($\alpha$-Si) metasurface and due to the geometry of utilized nano-pillar, non-polarizing operation is also achieved. With our proposal, it should be mentioned that the propagating direction and transverse mode profiles of the input and output beams are maintained, which would be convenient for cascading and cooperating with other element in optical systems. Specifically, the operation mechanism of the proposed MPBS is based on a complex grating-like diffraction phase pattern optimized by a gradient-descent-based iterative algorithm. As a proof of concept, we have applied the optimized phase-grating paradigm on elaborately designed $\alpha$-Si metasurfaces and fabricated different MPBS samples with various port number ($N$=2~7), power ratio and spatial distributions. To evaluate the performance, two parameters of total splitting efficiency (*TSE*) and beam-splitting ratio *similarity* are employed to characterize the efficiency and fidelity of fabricated samples. According to the experimental results, the *TSE* and *similarity* are 74.7%~80.7% and 78.4%~89.3% within the wavelength range of 1500~1600 nm, respectively. We believe that the proposed MPBS exhibits the powerful capabilities of metasurface on wavefront shaping and optical steering.

**RESULTS**

The functionality of our proposed free-space MPBS is to split the incident single wavelength light beam into multiple sub-beams with respectively predetermined power (denoted as normalized $P_1$~$P_N$) while the operation wavelength and polarization keep constant. With our proposal, an arbitrary MPBS can be readily achieved once the port number, splitting ratio as well as the spatial distributions of the output beams are specified. As a concrete example, **Figure 1a** shows the schematic of a 5-port beam splitter, in which the input light beam illuminates on the MPBS and splits into five sub-beams with predetermined power ratio of $P_1:P_2:P_3:P_4:P_5=1:2:3:4:5$. After propagating a certain distance along the respectively distinct deflecting direction, the sub-beams are received on the observation plane and defined as the output ports of the device. Here, for the specific case in **Figure 1b**, the spatial distribution of the output beams is considered as equally spaced on a circle of the received plane.

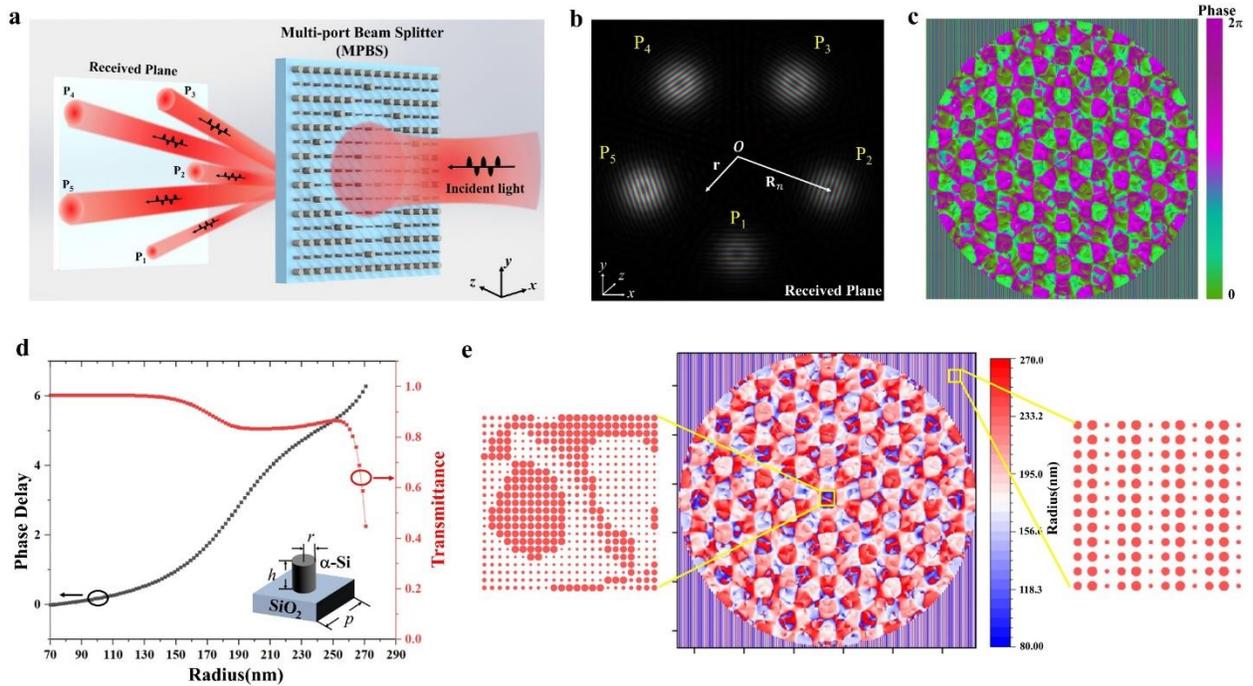

**Figure 1.** The schemes and design process for the metasurface-based free-space optical MPBS. (a) The schematic of a 5-port beam splitter example with power ratio of $P_1:P_2:P_3:P_4:P_5$. (b) The desired output on the received plane of the 5-port beam splitter with power ratio of $P_1:P_2:P_3:P_4:P_5=1:2:3:4:5$. (c)The optimized phase-only pattern obtained by gradient-descent-based algorithm for the 5-port beam splitting with power ratio of 1:2:3:4:5. (d) The simulated phase delay and transmittance versus the varied radius (70~270 nm) of $\alpha$-Si nano-pillars. Through systematically sweeping the parameters, the carefully selected height and period of the nanopillars are h=750 nm and p=630 nm, respectively. (e) The whole and partial arrangement diagrams of nanopillar radius for the metasurface-based MPBS with power ratio of 1:2:3:4:5.

A diffraction grating is the simplest structure to perform beam splitting. However, a single grating cannot split the incident beam into multiple output beams with non-uniform power ratio. Actually, a series of gratings are required to perform the task of multi-port beam splitting. Thanks to the distinctive wavefront control capability, a two-dimensional complex grating-like diffraction phase pattern can be achieved by the phase-gradient metasurface. Hence, the process to design a metasurface-based MPBS can be considered as two steps. The first step is to design a proper diffraction phase pattern to achieve the desired multi-port beam splitting and the second step is to determine the distribution of "meta-atoms" corresponding to the phase pattern acquired in first step. In the following text, we will consider a 5-port beam splitter shown in Figure 1b as a specific example to introduce the design process in detail.

Firstly, both the incident and output beam are considered as the same transverse mode profile. Theoretically, the MPBS is independent of the mode of both the input and output. In this work, the mode profile is specifically set as fundamental Gaussian mode since it is commonly utilized in most spatial optical systems. Besides, in our design, the number and the spatial distribution of the output ports can be flexibly settled. The only restriction is to avoid the overlap between adjacent output beams, which is corresponding to the divergence angle of the Gaussian beam, the angle of beam deflecting and the distance between the incident and the observation plane. Here, the case shown in Figure 1b is set as a specific circumstance. The incident beam, which is a Gaussian-mode with a beam waist radius of 40 μm, falls precisely on the MPBS and then splits into five sub-beams with the same deflecting angle of 4.1°. After a propagation distance of 6 mm, the output beams are received by an observation plane, representing uniformly distributed around a 430-μm-radius circle. Both the input and output light beams are at the wavelength of 1550 nm. Following these settled parameters, a diffraction phase pattern could be determined.

Before focusing on the concrete example, we should first determine how to design a diffraction phase pattern to deflect the incident wave to distinct outgoing directions at the same time. A straightforward but effective recipe is to utilize superposed phase gratings. It has been recently demonstrated that a mixed phase pattern will simultaneously perform respective phase compensation according to the superposition theorem.[49] As shown in Figure 1b, for convenience, the center of the incident light beam is considered as the origin of the coordinate and then the diffraction pattern composed of a series of blazed gratings can be expressed as:

$$F_1(\mathbf{r}) = \arg\{\sum_{n=1}^{N} A_n \exp\left[-i\mathbf{k}_n \cdot (\mathbf{r} - \mathbf{R}_n)\right]\} \tag{1}$$

where $N$ refers to the number of the output ports, and $A_n$ indicates the normalized amplitude of the $n_{th}$ sub-beam. Displacement vector $\mathbf{r}$ and $\mathbf{R}_n$ are the position coordinates of a certain point and the central point of the $n_{th}$ output beam on the transverse plane, respectively. $\mathbf{k}^n = (k_x^n, k_y^n)$ represents the corresponding transverse wave vector compensation of the $n_{th}$ sub-beam deflected from the incident beam, resulting in the deflection angle of $tan^{-1}(k_x^n/k)$ and $tan^{-1}(k_y^n/k)$ along $x$-axis and $y$-axis, respectively. Besides, considering the circular symmetric distribution on the transverse plane of the Gaussian beam, the functional area of diffraction pattern contributing by Equation 1 is purposely set as a circular region to match the incident beam. The Equation 1 is rewritten as:

$$F_2(\mathbf{r}) = F_1(\mathbf{r}) \times \chi(\mathbf{r}), \quad \chi(\mathbf{r}) = \begin{cases} 1, & |\mathbf{r}| < R_{threshold} \\ 0, & otherwise \end{cases} \tag{2}$$

where, $R_{threshold}$ is determined by the beam waist of the incident Gaussian spot.

Additionally, to spatially filter out the undesired light, a three-step diffraction grating is additionally implemented in the surrounding area, which is given by $F_{grating}(\boldsymbol{r}) = \exp(i\boldsymbol{k}_{grating} \cdot \boldsymbol{r})$.

Thus, the ultimate diffraction phase pattern is expressed as:

$$F(\boldsymbol{r}) = \arg\{\sum_{n=1}^{N} A_n \exp[-i\boldsymbol{k}_n \cdot (\boldsymbol{r} - \boldsymbol{R}_n)]\} \times \chi(\boldsymbol{r}) + \exp(i\boldsymbol{k}_{grating} \cdot \boldsymbol{r}) \quad (3)$$

However, the phase-only modulation induced by phase gratings cannot accurately accomplish the arbitrary power ratio beam splitting functionality, which involves both the amplitude and phase information. Thus, an iterative algorithm with gradient descent is employed to obtain the optimal phase-only pattern, which produces the light field distribution on the target plane to approximate an ideal distribution. The main procedure to generate the optimized phase pattern and the light field simulation method based on Discrete Fourier Transform (DFT) can be found in Supporting Information S1. **Figure 1c** is the finally optimized two-dimensional diffraction pattern to perform the 5-port beam splitting with power ratio of 1:2:3:4:5, which is comprised of the complex phase pattern in the circular area to generate target light field and the stripe pattern in the surrounding area to filter out the undesired beams. Compared with the diffraction pattern given by Equation 3, the optimized one contains more meticulous details, promising more accurate target light field (Supporting Information Figure S3). It should be mentioned that our utilized iterative algorithm with gradient descent is a backward optimization algorithm. Thus it is different from the greyscale computer-generated holography (CGH) algorithm, which is a kind of forward simulation method. The detailed discussions are included in Supporting Information S1. Besides, we have deduced the complex-number form of this optimization. This complex algorithm can make more effective use of the computing library than the real number algorithm, so it can obtain the results with much faster speed. Moreover, the optimization complexity of our utilized algorithm is independent on the number of ports, but only depends on the area and resolution of the optimization region (i.e., the total number of pixels). In this work, the optimization is running with MATLAB and a GTX 1660Ti GPU is used, and the resolution is 2048 by 2048. It takes just a few seconds to optimize a hologram.

After obtaining the phase pattern, we further apply the optimized phase-grating paradigm on an elaborately designed metasurface. Here, the metasurface is considered as $\alpha$-Si nano-pillar array on quartz substrate to achieve polarization independent operation. For more accuracy, we firstly measured the dielectric function of $\alpha$-Si sample deposited on 300 μm quartz substrate by plasma enhanced chemical vapor deposition (PECVD) by spectroscopic ellipsometry. The details are provided in Supporting Information S2. Typically, with the measured refractive index ($n$=3.324) and absorption coefficient ($k$=0.015) at the wavelength of 1550 nm, the phase delay and transmittance have been calculated by Finite-Difference-Time-Domain (FDTD) method within the radius range of 70~270 nm. The lattice constant and height of the nano-pillars have been carefully selected so that the phase delay can be nearly regulated from 0 to $2\pi$ and high transmission efficiency can be maintained simultaneously (**Figure 1d**). With the relation between the phase delay and the radius of the nano-pillars, the proper arrangement of nano-pillar radius (**Figure 1e**) can be obtained to construct the MPBS.

To verify our design shown in Figure 1e, some FDTD simulations have been carried out. Actually, due to the limited computing capacity of the simulations, the MPBS has a more compact size of 90 × 90 μm², i.e., 143 × 143 $\alpha$-Si nano-pillars involved in the simulation, and it is illuminated by an *x*-polarization Gaussian incident light with the radius of beam waist as 40 μm. **Figure 2a** shows the simulation result of the MPBS with predesigned power ratio of 1:2:3:4:5. **Figure 2b** shows the power distribution versus the azimuth angle $\theta$ and the peak power ratio of each sub-beam. It can be seen that the beam splitting ratio is quite consistent with the theoretical design.

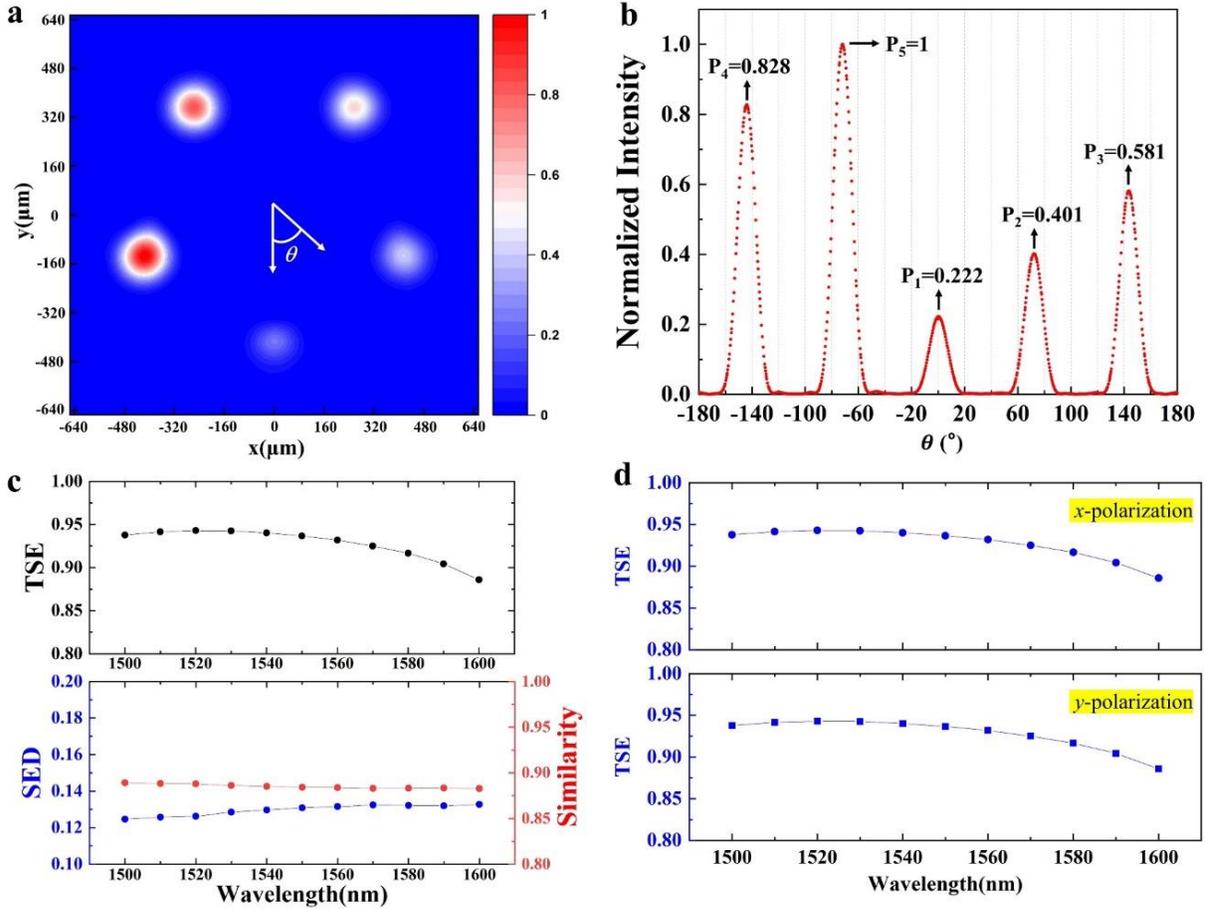

**Figure 2.** Simulation results of the proposed MPBS with power ratio of 1:2:3:4:5. (a) The output beams with normalized light intensity on the received plane. (b) The extracted normalized peak power ratio of each sub-beams versus the azimuth angle $\theta$, $\theta$ is defined as the polar angle with negative y-axis on the received plane, as shown in (a). (c) The calculated *TSE* and *SED* (*similarity*) in the wavelength range of 1500~1600 nm with wavelength interval of 10 nm. (d) The *TSE* for the *x*-polarization and *y*-polarization incident light.

Additionally, to evaluate the performance of the MPBS, two parameters are employed. The first one is the total splitting efficiency (*TSE*), which is the ratio of the total intensity of the desired sub-beams to the total intensity of all transmitted light and expressed by Equation 4. The other one is standardized Euclidean distance (*SED*),[50] which evaluates the discrepancy of the theoretical splitting ratio vector and the experimental one. In addition, *similarity* is also introduced to characterize the beam-splitting fidelity more clearly. The relation between *SED* and *similarity* is expressed as Equation 5. As shown in **Figure 2c**, the simulated *TSE* can reach >88% with *similarity* >88.2% (*SED*<0.133) in the wavelength range of 1500~1600 nm, which covers the C-band and most of L-band. Besides, to demonstrate the polarization independence of the MPBS, we have repeated the simulations with *y*-polarization Gaussian incident light and plot the results in **Figure 2d**. According to the simulation results of both *x* and *y*-polarization, our purposed MPBS is insensitive to the polarization states of the incident light beam.

$$TSE = \frac{\sum_{i=1}^{N} I_i}{I_{total}} \quad (4)$$

$$Similarity = \frac{1}{1+|SED|} \quad (5)$$

**EXPERIMENTAL RESULTS**

Further, several MPBS samples with various splitting ratio have been fabricated and measured. At first, $\alpha$-Si (thickness of 750 nm) was deposited on a quartz substrate by PECVD. Then, the MPBS samples were fabricated on $\alpha$-Si layer with electron beam lithography (EBL) and inductively coupled plasma reactive ion etching (ICP-RIE). It is worth mentioning that a hard mask of Cr layer was adopted to increase the aspect ratio of the nano-pillars, and an additional $SiO_2$ mask was also employed to avoid experimentally cumbersome lift-off process of Cr. The details of fabrication processes can be found in Supporting Information S3. **Figure 3** shows the optical microscope and scanning electron microscope (SEM) images of the fabricated sample with splitting ratio of $P_1:P_2:P_3:P_4:P_5=1:2:3:4:5$. All samples (five types of MPBSs) are comprised of 500×500 nano-pillars with size of 315×315 μm$^2$, which is comparable to the diameter of the incident light spot (~300 μm) located on the MPBS bracket. Actually, such sample size is determined as a trade-off between the performance and cost. Specifically, to achieve precise beam-splitting, large area is desired so that more meta-atoms could be implemented to achieve high resolution phase pattern. However, large area sample also introduce higher cost in terms of time cost of optimizing the phase pattern and particularly the expenditure of electron beam lithography. If the standard CMOS fabrication is adopted for future mass production, the cost of implementing large area MPBS could be significantly reduced.

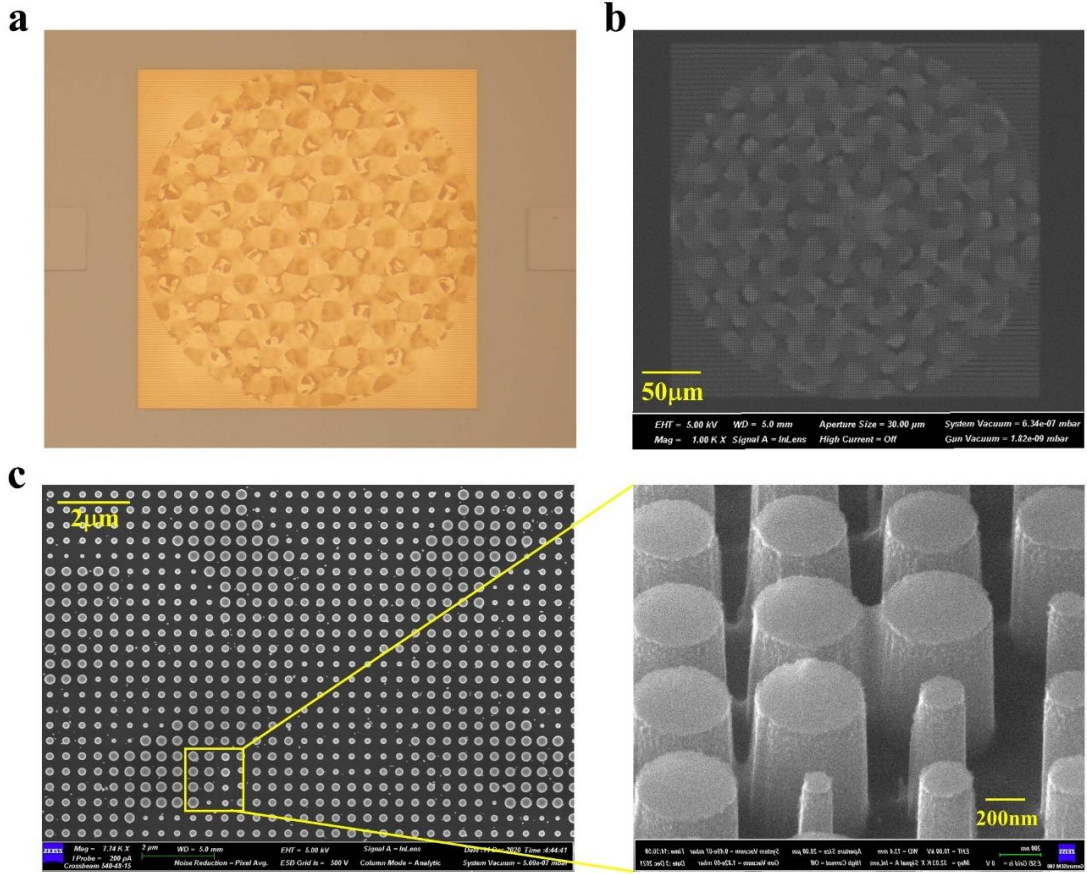

**Figure 3.** The optical microscope image and SEM images of the fabricated sample with splitting ratio of $P_1:P_2:P_3:P_4:P_5=1:2:3:4:5$. (a) The optical microscope image of the MPBS sample. (b) The overall SEM image of the MPBS sample. (c) The partially enlarged SEM images of the MPBS sample.

To measure the transmission of the fabricated samples, a confocal microscope system has been built up (more details are provided in Supporting Information S4). The inset image in **Figure 4a** is captured by CCD and shows a typical output of a 5-port sample with splitting ratio of 1:2:3:4:5 at the wavelength of 1550 nm. It can be seen that there is a central spot, which is the directly transmitted light without modulation and indicates that the diffraction efficiency of fabricated sample is lower than the design. The reason could be attributed to the sidewall roughness and diameter deviation of fabricated nano-pillar (detailed discussions can be found in Supporting Information S5). With the recorded data of CCD, the power distribution versus the azimuth angle $\theta$ can be extracted and shown in Figure 4a as red dots while the simulation results are also plotted as blue square for comparison. All results are normalized with the intensity of the maximum output beam. It can be found that the experimental results and simulations are quite consistent. **Figure 4b** and **4c** show both the *TSE* and *SED* (*similarity*) within the operation wavelength range of 1500~1600 nm, respectively. In both figures, the results of experiment/simulation are shown as red dots/blue squares. According to the experimental results, the *TSE* is 74.7%~78.8% and *similarity* is 78.4%~85% (*SED*: 0.177~0.276) within the bandwidth of 100 nm, respectively. These results indicate that our proposed MPBS could operate within a very broad bandwidth. The *TSE* of our experiments is reasonable in the previously reported transmissive meta-surface devices (discussions can be found in SI). The polarization-independent operation of the MPBS is also measured and shown in Figure S12 of SI. Furthermore, with our proposed MPBS, the incident beam can be guided towards any direction/angle. Thus, another case with a rhombic pattern of output sub-beam spots is shown in **Figure 4d~f** with power ratio of 1:2:3:4. The experimental *TSE* is 78%~79.4% and *similarity* is 85%~88.7% (*SED*: 0.128~0.177). In addition, all the experimental results of other samples are provided in Supporting Information S8 and summarized in Table 1.

From Table 1, it can be seen that different types of free-space MPBSs, including various beam-splitting number (up to 7) and spatial distribution of output ports (circle-arranged or rhombus-arranged), have be successfully achieved.

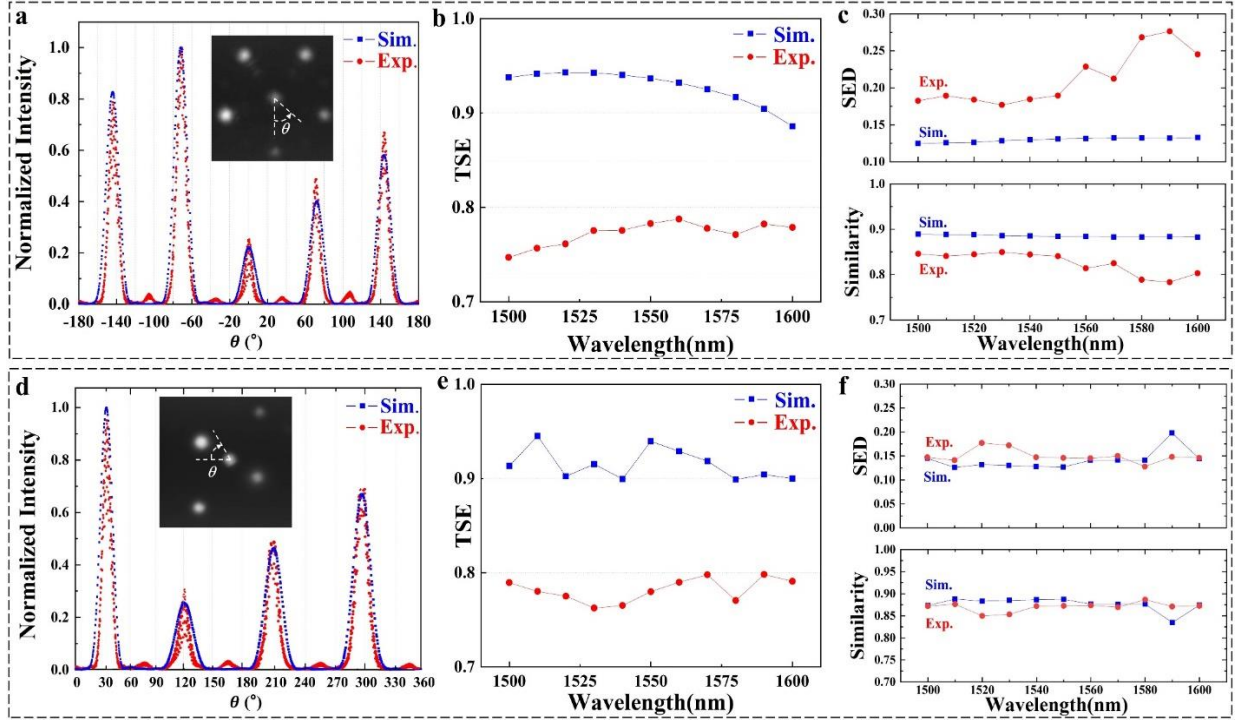

**Figure 4.** Experimental results of the proposed metasurface-based MPBSs with power ratio of 1:2:3:4:5 (a,b,c) and 1:2:3:4 (d,e,f). (a,d) The extracted normalized peak power ratio of each sub-beams versus the azimuth angle $\theta$, the inset shows the recorded CCD image. (b,e) The simulated and experimental *TSE* in the wavelength range of 1500~1600 nm with wavelength interval of 10 nm. (c,f) The simulated and experimental *SED/similarity*.

**Table 1.** The complete experimental results of all fabricated samples

| Beam-Splitting Ratio | *Exp.TSE* | *Exp.SED* | *Exp.similarity* |
|---|---|---|---|
| **1:2** | 76.4%~80.7% | 0.13~0.175 | 85.1%~88.5% |
| **1:2:3:4** | 78%~79.4% | 0.128~0.177 | 85%~88.7% |
| **1:2:3:4:5** | 74.7%~78.8% | 0.177~0.276 | 78.4%~85% |
| **1:1:1:1:1** | 74.7%~78.1% | 0.16~0.243 | 80.5%~86.2% |
| **1:1:1:1:1:1:1** | 77.1%~79.5% | 0.12~0.175 | 85.1%~89.3% |

**CONCLUSION**

In summary, we have proposed and demonstrated a free-space optical MPBS based on a polarization-independent all-dielectric metasurface. An iterative algorithm with gradient descent optimization is utilized to design the phase grating for precisely multi-beam splitting. With such a generic optimization paradigm, a series of MPBSs have been designed and fabricated on $\alpha$-Si substrate with various port number ($N$=2~7), splitting ratio and spatial distributions. According to the complete experimental results of all fabricated samples, the calculated *TSE* and *similarity* are 74.7%~80.7% and 78.4%~89.3% within the bandwidth of 100 nm. Compared with previous metasurface-based 2-port BSs in visible/near-infrared regimes,[44]-[48] our proposed MPBS not only demonstrates multi-port beam-splitting successfully, but also achieves arbitrary power ratio along with arbitrarily predetermined port number and the spatial distribution of output beams. This work further exhibits the flexibility of metasurface to achieve the functionalities that are hard to be obtained by conventional optical devices. Moreover, for the proposed MPBS, the characterizations of maintaining propagating directions and transverse mode profile of the incident beam are very beneficial for cascading and cooperating in expanded optical systems. We believe that such MPBS device could be promising for applications of LiDAR, beamforming networks, quantum optics and optical computing.

## ASSOCIATED CONTENT

**Supporting Information**

Additional explanation for (1) simulation and optimization method of diffraction phase pattern; (2) experimental measurement for the dielectric function of the $\alpha$-Si sample; (3) fabrication process of metasurface-based MPBS; (4) the experimental platform for characterizing the fabricated MPBS; (5) the analysis of *TSE* difference between the experiments and simulations; (6) the diffraction efficiency comparison with previous reports; (7) the experimental polarization performance of MPBS sample; (8) the complete experimental results of all fabricated samples.

## AUTHOR INFORMATION

**Corresponding Author**

**Xue Feng -** *Department of Electronic Engineering, Tsinghua University, Beijing 100084, China;* orcid.org/0000-0002-9057-1549; E-mail: x-feng@tsinghua.edu.cn.

**Yidong Huang -** *Department of Electronic Engineering, Tsinghua University, Beijing 100084, China;* E-mail: yidonghuang@tsinghua.edu.cn.

**Author Contributions**

T.T. and X.F. conceived the idea. T.T. designed and performed the simulations, experiments and data analysis. Y.L. contributed significantly to the numerical simulations. K.C., F.L. and W.Z. provided useful discussions and comments. T.T. and X.F. wrote the paper. Y.H. revised the manuscript. The manuscript was written through contributions of all authors. All authors have given approval to the final version of the manuscript.

**Funding Sources**

This work was supported by the National Key Research and Development Program of China (2018YFB2200402), the National Natural Science Foundation of China (Grant No. 61875101)

**Notes**

The authors declare no competing financial interest.

ACKNOWLEDGMENTS

This research was supported by the National Key Research and Development Program of China (2018YFB2200402), the National Natural Science Foundation of China (Grant No. 61875101), This work was also supported by Beijing academy of quantum information science, Beijing National Research Center for Information Science and Technology (BNRist), Frontier Science Center for Quantum Information, Tsinghua University Initiative Scientific Research Program and Huawei Technologies Co. Ltd.. The authors would like to thank Mr. Zhiyao Ma and Prof. Yongzhuo Li for their valuable discussions and helpful comments.

# Metasurface-Based Free-Space Multi-Port Beam Splitter with Arbitrary Power Ratio

Tian Tian[1], Yuxuan Liao[1], Xue Feng[1]*, Kaiyu Cui[1], Fang Liu[1], Wei Zhang[1], and Yidong Huang[1]*

[1]Department of Electronic Engineering, Tsinghua University, Beijing 100084, China

*Correspondence: x-feng@tsinghua.edu.cn; yidonghuang@tsinghua.edu.cn

## S1. Simulation and optimization method of diffraction phase pattern

### S1. 1. Light field simulation method based on Discrete Fourier Transform (DFT)

In this work, we have utilized discrete Fourier transform (DFT) to derive an accurate and fast light field simulation method, and successfully applied to the light field simulation of diffraction phase pattern to investigate the beam splitting result.

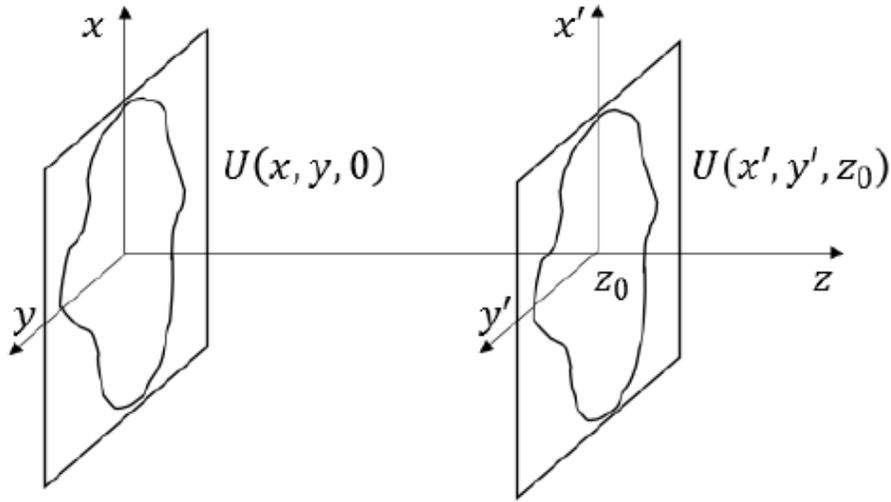

**Figure S1.** The input and output plane diagram of light field propagation problem

As depicted in Figure S1, for a monochromatic light (wavelength denoted as $\lambda$) propagation, the space distribution of its complex amplitude $U$ satisfies Helmholtz equation:

$$(\nabla^2 + k^2)U = 0, k = \frac{2\pi}{\lambda} \tag{S1}$$

Using Green's theorem to solve this equation, we can derive the Rayleigh-Sommerfeld integral, which is used to calculate the target light field propagating from a given incident plane to the diffraction plane[1]:

$$U(x',y',z_0) = -U(x,y,0) * \frac{1}{2\pi} \frac{\partial}{\partial z}\left[\frac{e^{-jkr}}{r}\right]\bigg|_{z=z_0} = U(x,y,0) * \frac{1}{2\pi}\left(\frac{e^{-jkr}}{r} \frac{z_0}{r}(jk + \frac{1}{r})\right)\bigg|_{z=z_0} \tag{S2}$$

where $r = \sqrt{x^2 + y^2 + z^2}$; $*$ refers to the convolution.

In order to be processed on a computer, we need to sample the input light field (*i.e.* discretization) and discrete Fourier transform (DFT) are utilized to rewrite the convolution formulated in eq S2:

$$U(p\Delta x, q\Delta y, z_0) = \sum_{r,s} U(r\Delta x, s\Delta y, 0) h((p-r)\Delta x, (q-s)\Delta y; z_0) \Delta x \Delta y \tag{S3}$$

Herein, we specify the sampling period of input optical field as $\Delta x$ and $\Delta y$, in $x$ and $y$ direction respectively. The sampling point counts are defined as $N, M$ and stored in matrix $U \in C^{M \times N}$, then $r \in [-N/2+1, N/2+1], s \in [-M/2+1, M/2+1]$.

## S1. 2. The main procedure to optimize the diffraction phase pattern

As described in the main text, a modulation function (eq S4) composed of a series of blazed gratings is required to perform the arbitrary power ratio beam splitting. Nevertheless, a diffraction phase grating, which only generates phase gradient, cannot accurately carry both the amplitude and phase information. To tackle such problem, an iterative algorithm is employed to obtain the optimal phase-only pattern.

$$F(\boldsymbol{r}) = \arg\{\sum_{n=1}^{N} A_n \exp[-i\boldsymbol{k}_n \cdot (\boldsymbol{r}-\boldsymbol{R}_n)]\} \times \chi(\boldsymbol{r}) + \exp(i\boldsymbol{k}_{grating} \cdot \boldsymbol{r}) \tag{S4}$$

The basic principle of optimizing this problem is gradient descent method. In our specific problem, it is significant to note that we need to differentiate real valued functions with complex valued variables to figure out the steepest descent direction. While, the conclusion of ref [2] reveals that under relatively loose conditions, we can also find a variable similar to gradient, which gives locally steepest descent direction, and the primitive gradient descent method can still be applied. Figure S2 depicts the theorem framework of the gradient-descent-based optimization algorithm. More specific mathematical derivation can be found in ref [3].

It should be noticed that our utilized iterative algorithm with gradient descent looks like a kind of greyscale computer-generated holography (CGH) algorithm. But they are different. Actually, our utilized iterative algorithm with gradient descent is a backward optimization algorithm while the CGH algorithm is a kind of forward simulation method. The CGH is a computational algorithm to generate holographic diffraction patterns that can be used to reconstruct a 3D image of an object. The algorithm uses mathematical models to simulate the diffraction of coherent light passing through the object and interference with a reference beam on a flat surface. It is a forward simulation. The gradient descent, on the other hand, is an optimization algorithm used in machine learning and deep learning to find the minimum value of a loss function. The algorithm starts with an initial guess of the solution and iteratively adjusts the solution in the direction of the negative gradient of the loss function until a minimum is found. The purpose of gradient descent is to optimize the parameters of a model to minimize the error between the predicted outputs and the actual outputs. In our work, we are designing a hologram that interacts with specific incident light modes and obtains a desired light field at distance. So, it is a backward optimization.

**Input**: Target phase grating: $P_t \in \mathbb{C}^{M \times N}$; guess phase grating: $P_0 \in \mathbb{C}^{M \times N}$; propagation distance: $z_0$
input light field: $U_I \in \mathbb{C}^{M \times N}$; other required hyperparameter sets: $\mathbb{R}_{h_0}$
Optimizer: $\mathcal{O}$; loss function: $l$; Number of Iterations: $T$

**Ouput**: phase-only diffraction grating $P_0$
- Initialize $\mathcal{O}$ and $l$
- **for** $i=1$ to $T$ **do**

  Compute $\frac{\partial}{\partial P_{i-1}^*} \mathcal{L}$

  $$P_i = \mathcal{O}.\text{step}\left(\frac{\partial}{\partial P_{i-1}^*} \mathcal{L}\right)$$

  **End for**
  Return $P_o = \frac{P_T}{|P_T|}$

**Figure S2.** The theorem framework of the gradient-descent-based optimization algorithm.

Then, we have applied this algorithm to the optimization of beam splitting phase pattern. Specifically, for the MPBS with power ratio of 1:2:3:4:5, Figure S3 shows the comparison of the original phase pattern (directly given by eq S4) and the optimized one, along with their output pattern obtained by DFT-based light field simulation method (S1). Comparing with the initial diffraction phase pattern, the optimized one contains more meticulous details, and promises more accurate target light field. In addition, the optimized phase patterns and simulated output beams of other type MPBSs with various port number ($N$=2~7), power ratio and spatial distributions are also provided in Figure S4. Here, the optimization algorithm is running with MATLAB and a GTX 1660Ti GPU is used, and the resolution is 2048 by 2048. It takes just a few seconds to optimize a hologram while the time cost of meta-atom designing have been excluded.

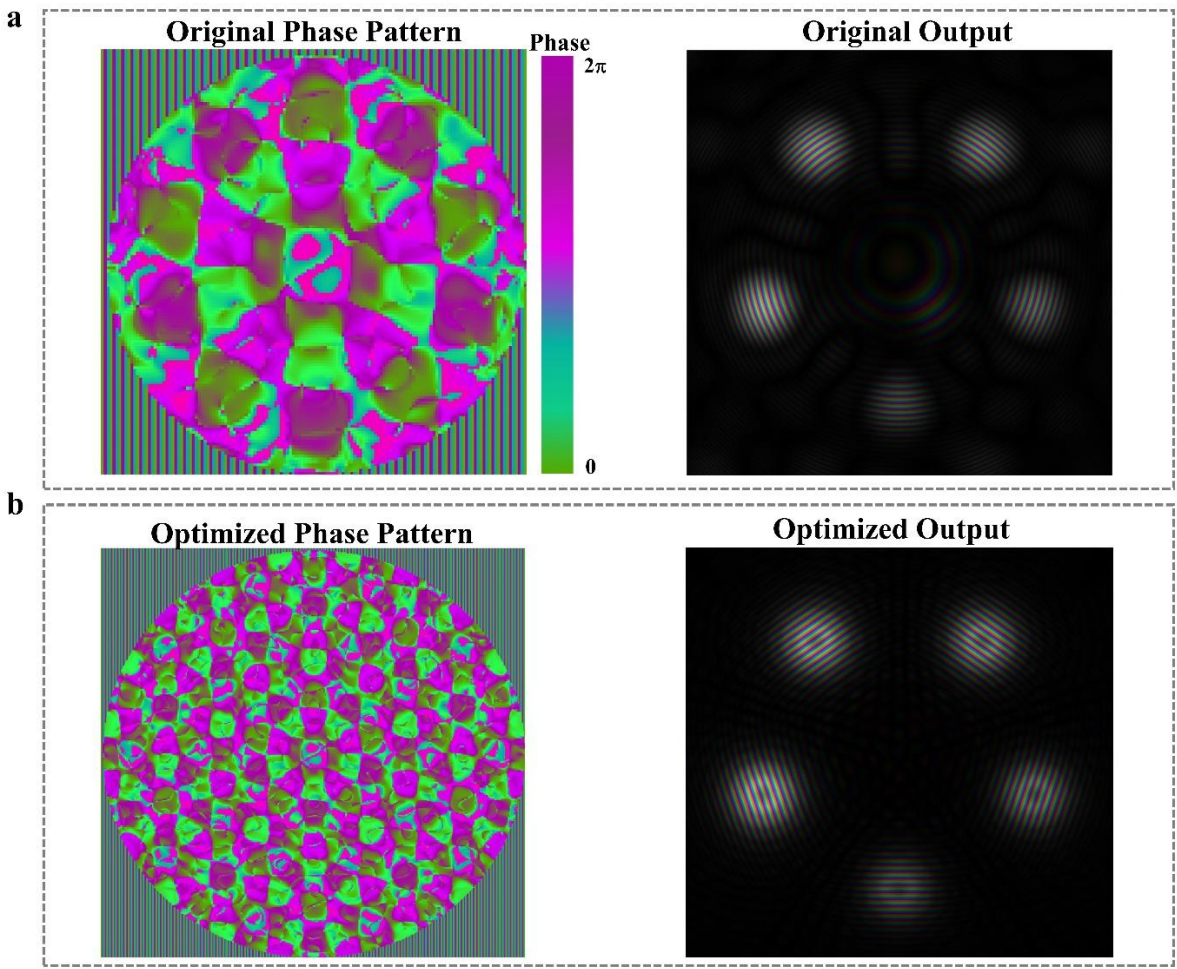

**Figure S3.** The comparison of the phase patterns and corresponding output beams before and after optimization. (a) The original phase pattern and output. (b) The optimized phase pattern and output.

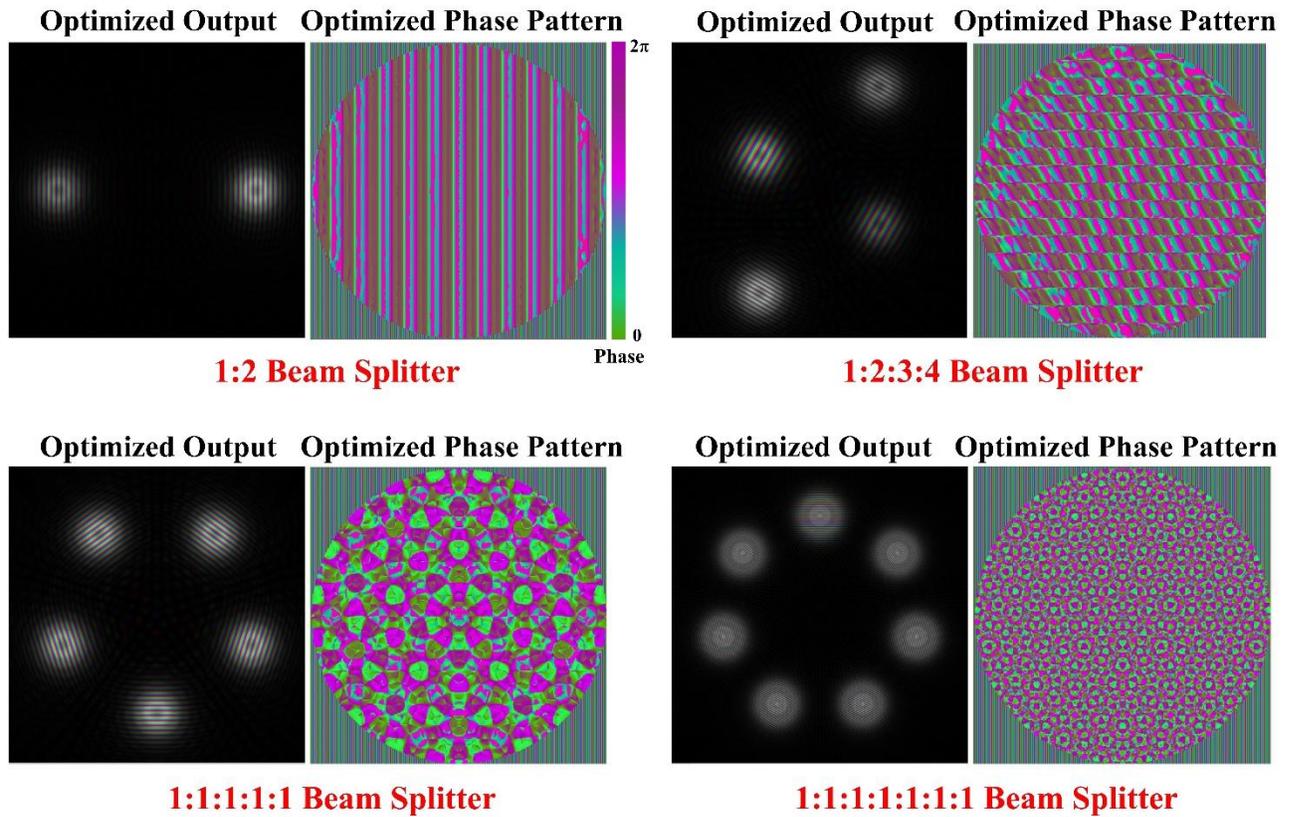

**Figure S4.** The optimized phase patterns and corresponding output beams of other type MPBSs. (a,b,c,d) The optimized phase pattern and simulated output for the MPBS with power ratio of 1:2 (a); 1:2:3:4 (b); 1:1:1:1:1 (c); 1:1:1:1:1:1:1 (d). The color of phase patterns indicates the phase values.

## S2. Experimental measurement for the dielectric function of the α-Si sample

In order to improve the simulation design and evaluate the actual device performance, we first measured the dielectric function of the α-Si sample on the quartz substrate. Plasma enhanced chemical vapor deposition (PECVD) was used to deposit 750nm-thick α-Si on a 300nm-thick quartz, which was beforehand sputtered on Si substrate. The spectroscopic ellipsometry (SENTECH SE850 NIR) was utilized to measure the refractive index ($n$) and absorption coefficient ($k$) at the wavelength of 700~2500nm, which is shown in Figure S5. It can be seen that the refractive index at the considered operation wavelength of 1550nm is 3.324, and the absorption coefficient is 0.015.

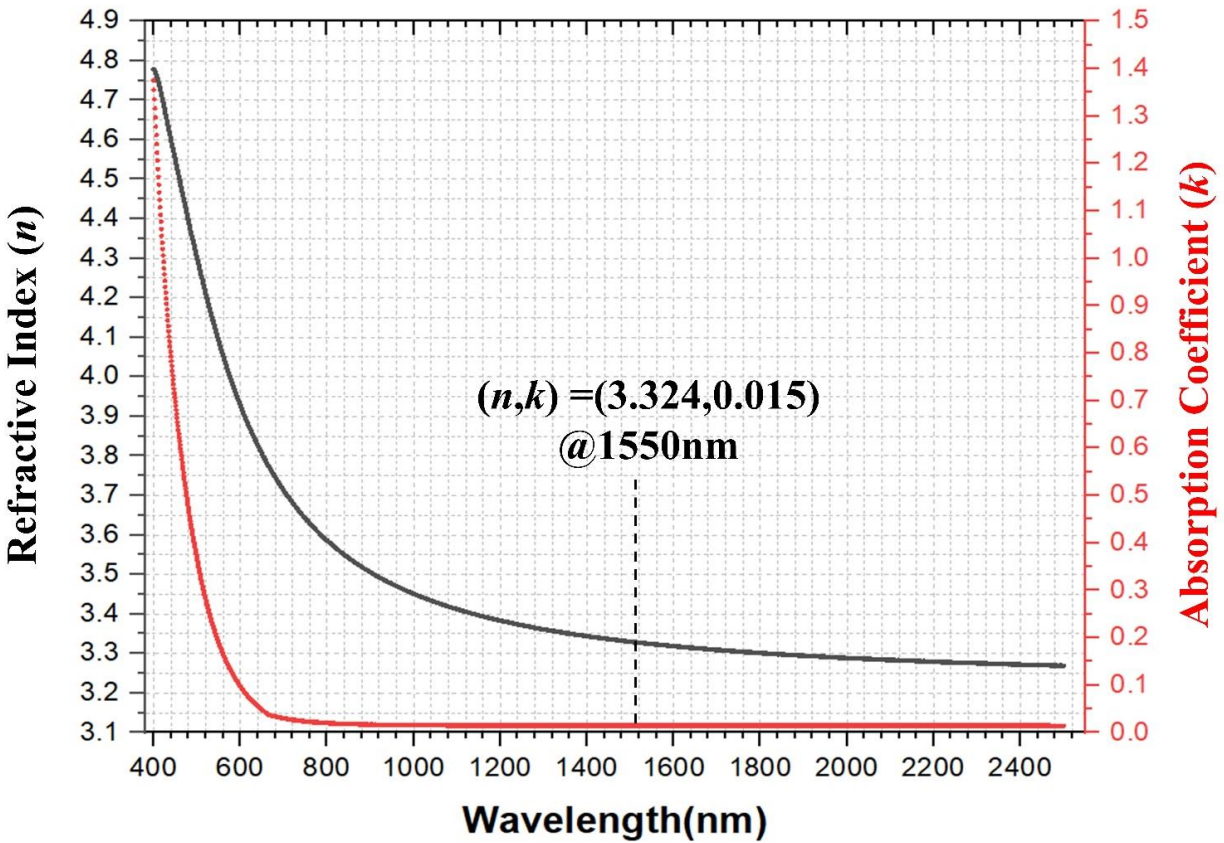

**Figure S5.** The experimentally measured dielectric function of the α-Si sample.

## S3. The fabrication process of metasurface-based MPBS

At first, a 750nn-thick layer of $\alpha$-Si is deposited on quartz substrate by plasma enhanced chemical vapor deposition (PECVD) and then a process of electron beam (EB) evaporation is utilized to deposit a 100nm-thick layer of Cr as metal hard mask in the following $\alpha$-Si etching process. Particularly, a 100nm-thick layer of $SiO_2$ is grown on the Cr layer as an additional hard mask to avoid experimentally cumbersome and uncontrollable lift-off process of Cr in the subsequent procedure. After that, a layer of negative photoresist (ZEON ZEP520A) is spun on top of the sample and then exposed by the electron beam lithography (EBL). Then, the pattern is transferred from the electron beam resist to the $SiO_2$ layer by the process of reaction ion etching (RIE) and electron beam resist removal, in the wake of which, inductively coupled plasma reaction ion etching (ICP-RIE) is utilized to etch Cr layer in the presence of $SiO_2$ mask. After the remaining $SiO_2$ is removed through buffered oxide etchant (BOE) corrosion, the designed pattern of beam splitter metasurface is finally implemented on $\alpha$-Si layer by ICP etching and Cr removal.

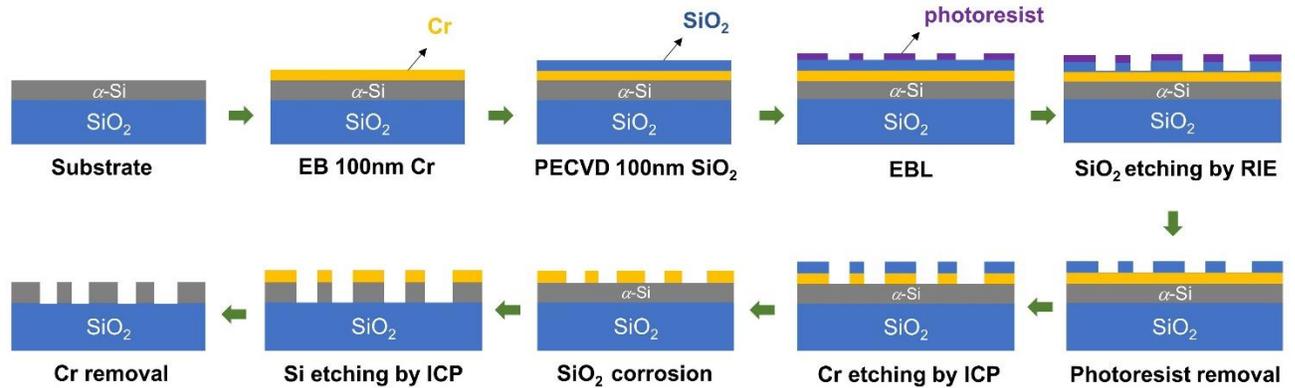

**Figure S6.** The fabrication process of the samples.

## S4. The experimental platform for characterizing the fabricated MPBS

To measure the transmission of the fabricated samples, a confocal microscope system has been built up (Figure S7). The input light source emits from a tunable laser (SANTEC TSL-710) with operation wavelength range of 1480~1640 nm so that the broadband operation can be experimentally featured. After a collimator, the Gaussian-mode input light is expanded by a two-stage lens and incident on the sample, which is vertically fixed on a precision displacement stage. After passing through the sample, the splitting beams would be focused on the CCD (ARTCAM-0016TNIR) by an objective lens (N10X-PH-Nikon) and recorded. In our setup, there is an additional near-infrared LED (M1450L3-LED) and a pellicle beam splitter to reversely illuminate the metal mark on the chip so that the sample can be located and aligned with the input laser beam.

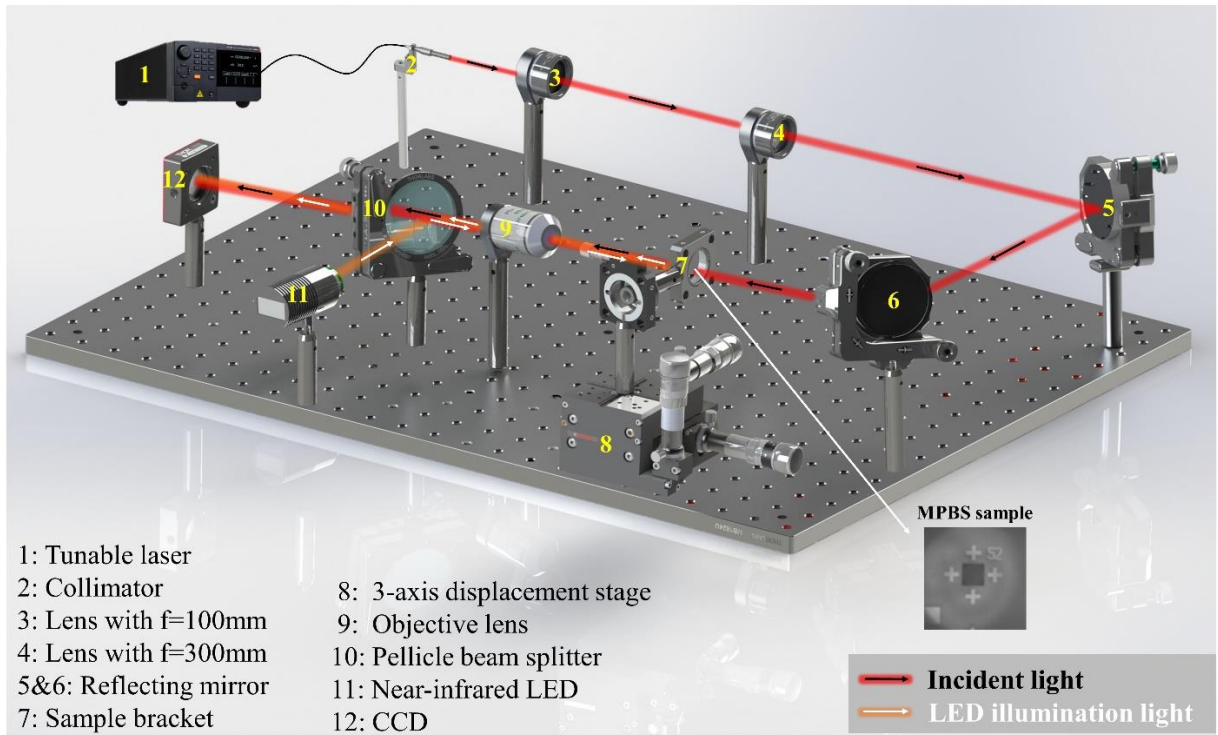

**Figure S7.** The experimental platform to characterize the MPBS samples.

## S5. The analysis of *TSE* difference between the experiments and simulations

In the process of fabricating metasurface-based MPBS samples, there are two aspects of geometric fluctuations of nano-pillars (meta-atoms). One is the rough sidewall shown in Figure S8a, which is inevitably introduced by etching process. The other one is the radial deviation from the designed value, which is mainly attributed to finite-precision of EBL. The actual diameter deviation can be characterized through SEM image (Figure S8b) and the maximal deviation value is about 14nm.

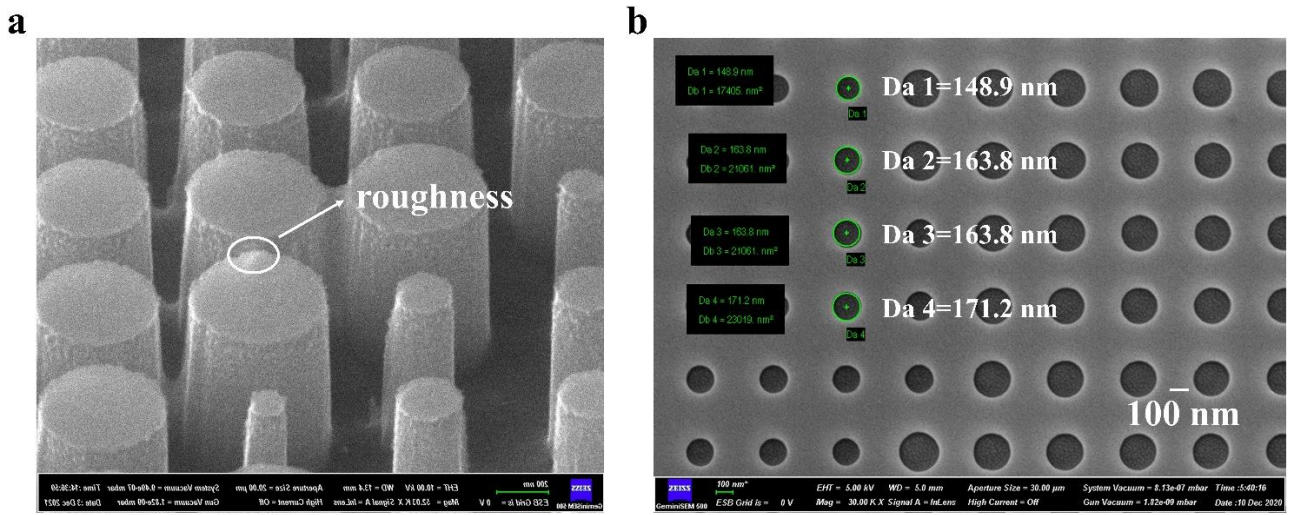

**Figure S8.** The two aspects of geometric fluctuations of nano-pillars. (a) The rough sidewall and (b) the radial deviation.

In order to quantitatively analyze the influence of the aforementioned two kinds of fabrication errors on experimental *TSE*, corresponding simulations are carried out by FDTD method. For the sidewall roughness, a simulated fluctuation with Gaussian distribution (mean value:0; standard deviation:3nm) is added to the geometric edge of each meta-atom to mimic the roughness (Figure S9). Figure S9b shows the simulated phase delay and transmittance versus varied radius (70~270 nm) of the smooth and rough nano-pillars. The result indicates that the sidewall roughness would introduce a decline in the transmittance at 1550 nm wavelength. The averaged deterioration is estimated as ~13.3%.

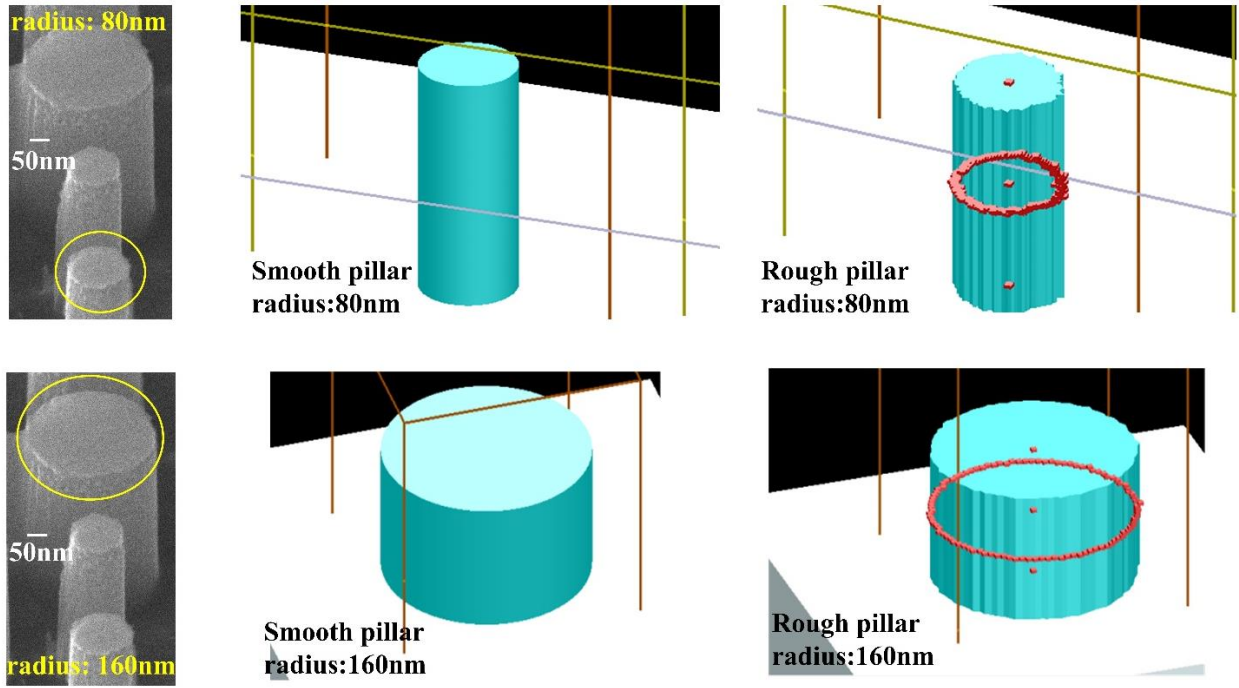

**Figure S9.** The simulation setup to mimic the rough sidewalls of fabricated nano-pillars.

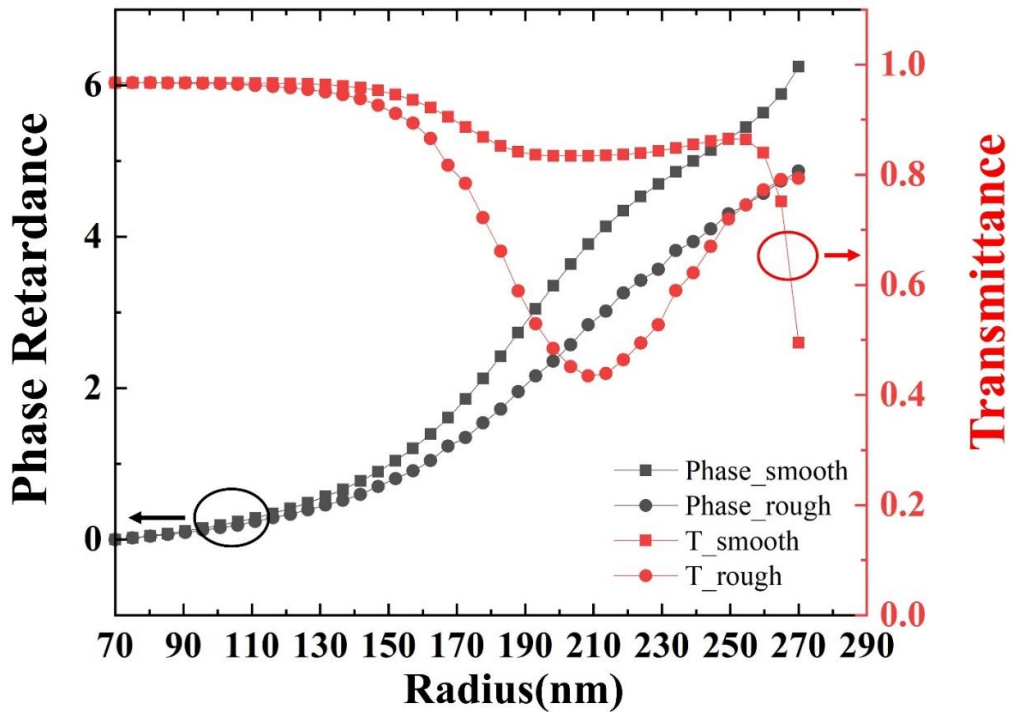

**Figure S10.** the simulated phase delay and transmittance versus the varied radius (70~270 nm) of the smooth and rough nano-pillars.

Besides the roughness, another simulation about the radial deviation is also carried out. With the MPBS with power ratio of 1:2:3:4:5 at the operating wavelength of 1550 nm, the radius deviation ($\Delta r$) with Gaussian distribution (mean value:0; standard deviation:2.5nm) is added to each meta-atom. Figure S11 shows the radius distributions and simulated output beam spots for the ideal and fluctuated case, respectively. It can be seen that in the presence of radius deviation, the central spot, which is the directly transmitted light without modulation, occupies about 4.2% of the total intensity.

According to the results shown in Figure.S10 and S.11, we believe the deterioration of *TSE* is attributed to fabrication errors in terms of sidewall roughness and radial deviation of the nano-pillar.

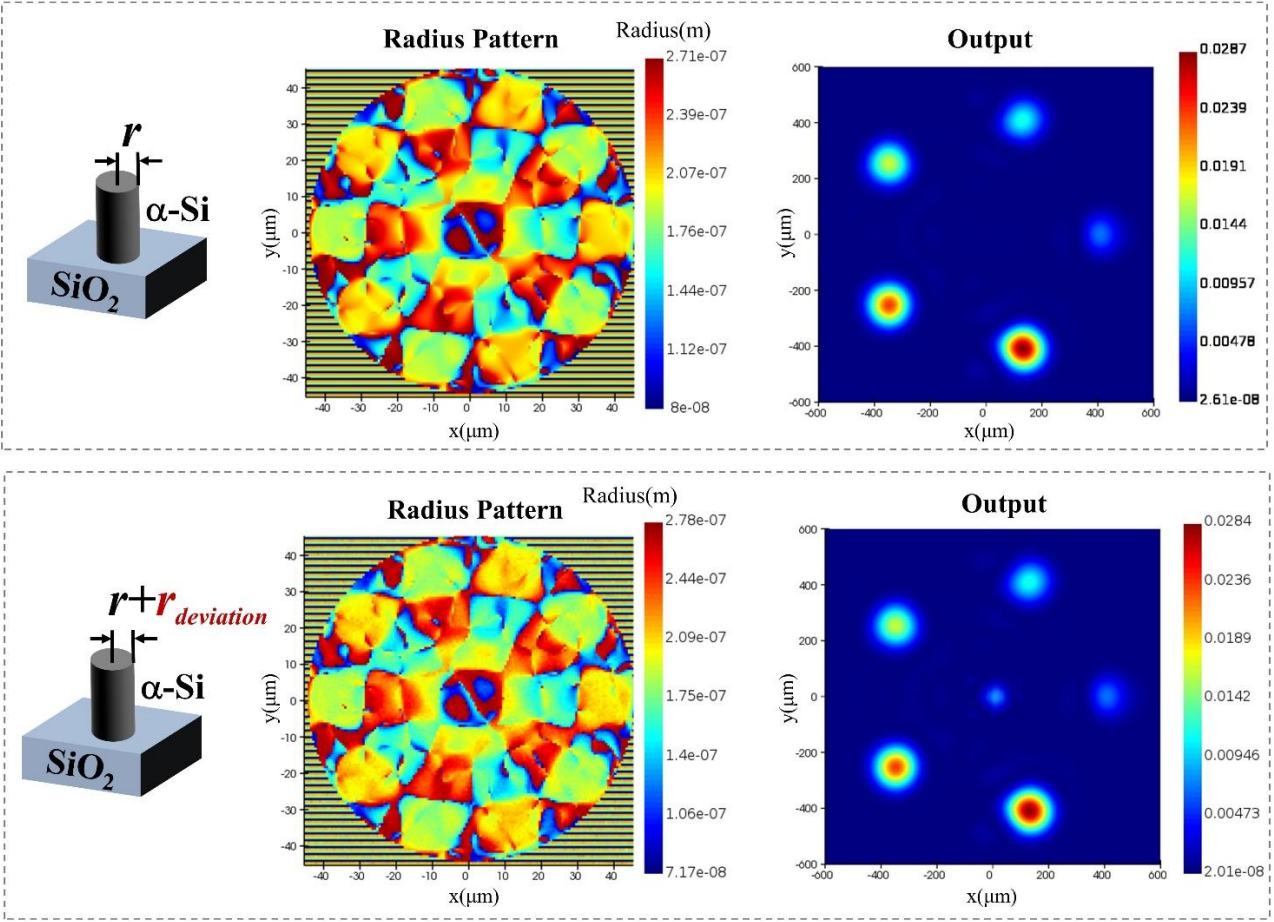

**Figure S11.** The radius distributions and simulated output results in ideal and fluctuated (i.e. with Guassian-distribution radius deviation) case.

## S6. The diffraction efficiency comparison with previous reports

Table S1 summarizes some representative data for the diffraction efficiency of previous reports, including reflective and transmissive metasurface-based BSs and some other transmissive devices such as the anomalous refractor, multi-wavelength BS, metalens. For the typical case of our proposed MPBS with power ratio of 1:2:3:4:5, the theoretical/simulated *TSE* is 93.6% and the experimental one is 78.3% (at wavelength of 1550 nm). Compared to previous reports, the *TSE* of our experiments is much better than the metal 2-port BSs, and is reasonable in terms of the transmissive metasurface-based devices including 2-port BSs and meta-lens. Besides, as discussed in S5, it is potential to improve the performance of our fabricated MPBS with reduced fabrication errors.

**Table S1.** Some representative data for the diffraction efficiency of previous reports

| Literature | Type | Diffraction efficiency |
|---|---|---|
| *ACS Photonics 2018, 5, 2997−3002* [4] | Reflective metasurface-based 2-port BS | 20.30% |
| *Nano Lett. 2022, 22, 5, 2059 2064* [5] | Reflective metasurface-based merging/2-port BS | Scheme A: 38%~58% Scheme B: 62%~75% (wavelength: 580~670nm) |
| *Nanomaterials 2021, 11, 1137* [6] | Transmissive metasurface-based 2-port BS | 93.21% (theory) |
| *ACS Photonics 2018, 5, 2402−2407* [7] | Transmissive metasurface-based anomalous refractor | 60%~80% for $\theta_{in} = +50°$ and $\theta_{out} = -50°$ (wavelength: 1125~1200nm) |
| *Adv. Optical Mater. 2017, 5, 1700645* [8] | Transmissive (silicon) multiwavelength metasurface | 60%~90% (wavelength: 1000~1300nm) |
| *Scientific Reports (2020) 10:7124* [9] | Transmissive metalens | 77% |

## S7. The experimental polarization performance of MPBS sample

In this work, nano-pillar array is employed for the sake of polarization independent operation. With the help of a polarization controller, the cross-polarization response can be experimentally investigated to feature the polarization dependence of the MPBS device, which is shown in Figure S12. The experimental results manifest that there is only a little difference in *TSE* response between *x* and *y*-polarization, which can be attributed to the imperfect fabrication of the circular nano-pillars.

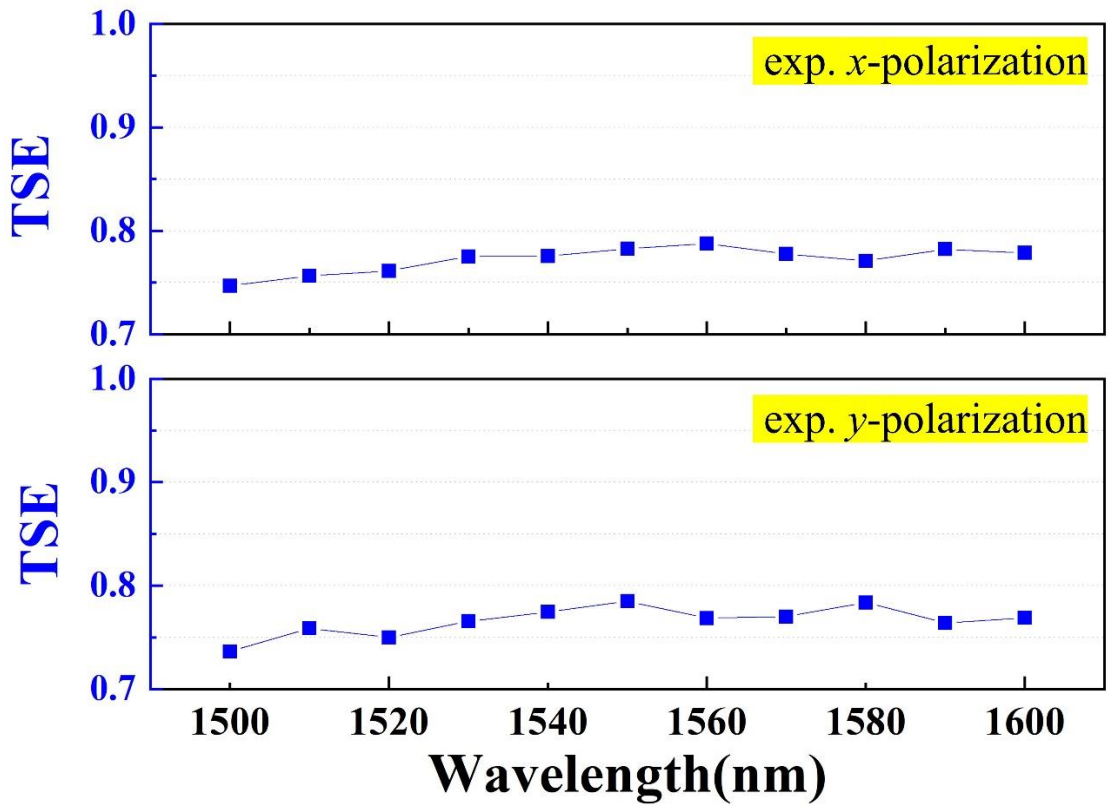

**Figure S12.** The *TSE* comparison plots for the original and crossed polarization incident light.

## S8. The complete experimental results of all fabricated samples

In order to present the generality of our design method, we have prepared five-type metasurface-based MPBSs and measured all of them. Besides the representative examples of 1:2:3:4:5 and 1:2:3:4 beam splitter in the main text, the complete experimental results of the other three samples are provided in this section, covering various cases of unequal power ratio and non-circular arrangement of sub-beams. The number of sub-beams prepared in our experiment is up to 7.

### S8. 1. Experimental results at 1550 nm

For the operating wavelength of 1550 nm, Figure S13 shows the experimental results with predesigned power ratio of 1:2; 1:1:1:1:1; 1:1:1:1:1:1:1, respectively. With the data recorded by CCD, the power distribution versus the azimuth angle $\theta$ can be extracted and shown as red dots while the simulation results are also plotted as blue squares for comparison. Both the experiment and simulation results are normalized with the intensity of the maximum output beam.

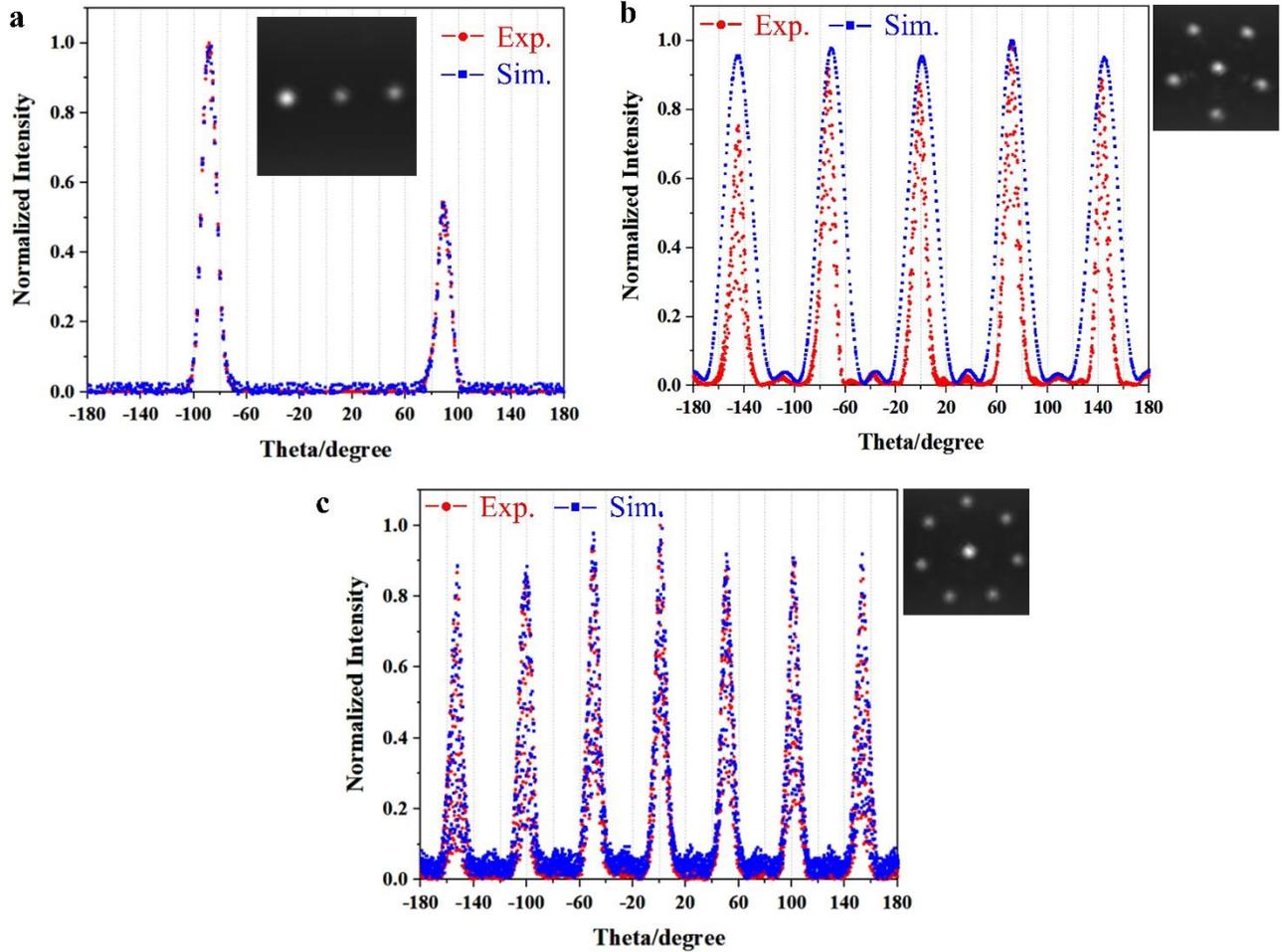

**Figure S13.** The experimental results of the other three samples at operating wavelength of 1550 nm. (a,b,c) The extracted normalized peak power ratio of each sub beams versus the azimuth angle θ for the MPBS with power ratio of 1:2 (a); 1:1:1:1:1 (b); 1:1:1:1:1:1:1 (c).

## S8. 2. Experimental Results within 100 nm bandwidth

Figure S14 shows the experimental *TSE* and *similarity* (*SED*) of another three samples within the operation wavelength range of 1500~1600 nm. Both the *TSE* and *similarity* (*SED*) are calculated by the values recorded by CCD and the formulae of eq 4 and eq 5 in the main text. In all figures, the experimental results are shown as red dots while the simulation ones are blue squares. According to the experimental results, the *TSE* is 74.7%~80.7% and *similarity* is 78.4%~89.3% (*SED*: 0.12~0.276) within the bandwidth of 100 nm, respectively.

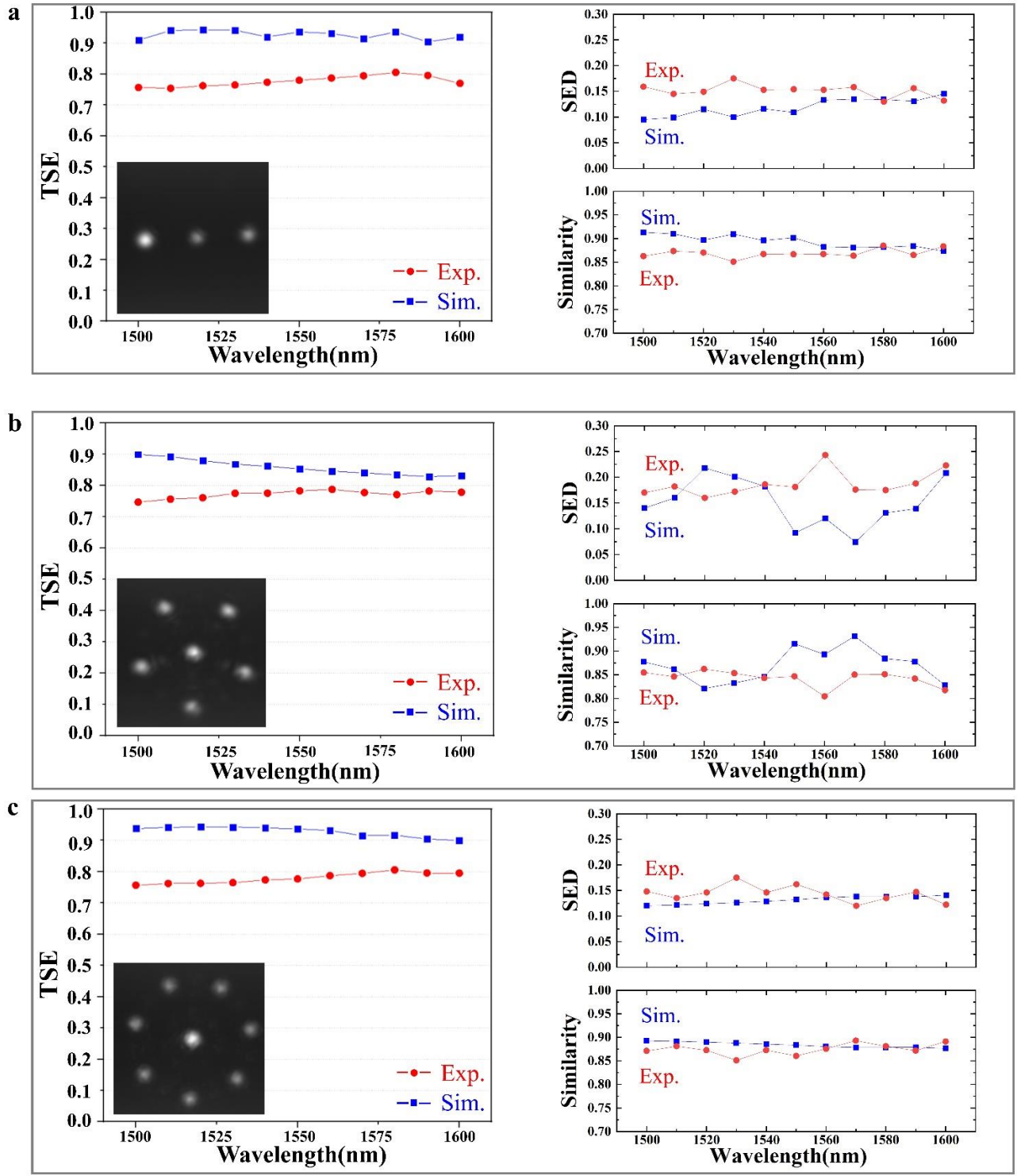

**Figure S14.** The simulated and experimental *TSE* and *similarity* (*SED*) of the other three samples

within the operation wavelength range of 1500~1600 nm. The *TSE* and *similarity* (*SED*) are shown with wavelength interval of 10 nm for the MPBS with power ratio of 1:2 (a); 1:1:1:1:1 (b); 1:1:1:1:1:1:1 (c).